\documentclass[aps,11pt]{revtex4}
\usepackage[dvips]{graphicx}

\usepackage{amsmath}
\usepackage{bm}

\begin{document}

\preprint{}

\preprint{}
\title{UPPER BOUND ON THE PROTON LIFETIME AND THE MINIMAL NON-SUSY 
GRAND UNIFIED THEORY\footnote{\bf{Based on the Talks given at NURT'06 
(V International Symposium on Nuclear and Related Techniques, 
Havana, CUBA, 3-7 April 2006), PLANCK'06 
(The ninth European meeting "From the Planck scale to the 
ElectroWeak scale'', Paris, FRANCE, 29 May 29 - 2 June,2006) 
and SUSY'06 (14th International Conference on Supersymmetry 
and the Unification of Fundamental Interactions. Irvine, 
California, USA. 12-17 June 2006).}}}
\author{Pavel Fileviez P\'erez}
\email{fileviez@cftp.ist.utl.pt}
\affiliation{Centro de F{\'\i}sica Te\'orica 
de Part{\'\i}culas.\ Departamento de
F{\'\i}sica.\ Instituto Superior T\'ecnico.\ Avenida Rovisco Pais,
1\\ 1049-001 Lisboa, Portugal}
\homepage{http://cftp.ist.utl.pt/~fileviez}

\begin{abstract}
In this talk we show that it is possible to find an upper bound 
on the total proton lifetime. We conclude that the minimal realistic 
non-supersymmetric grand unified theory is the modified 
Georgi-Glashow model with a Higgs sector composed of 
$5_H$, $24_H$, and $15_H$. We discuss the possibility to test 
this scenario at the next generation of proton decay experiments and 
future colliders through the production of light scalar leptoquarks.     
\end{abstract}
\maketitle

\section{Introduction}

It is well-known that the decay of the proton is the most 
dramatic prediction coming from matter unification~\cite{PatiSalam}. For this 
reason we believe that the best way to test the grand 
unified theories is through proton decay. 

In the first part of my talk I will discuss the 
possibility to find an upper bound on the proton lifetime, 
which is crucial to understand the possibility to test 
the different minimal grand unified models with the standard 
model matter content. 

As we know the minimal Georgi-Glashow model~\cite{GG} is the 
simplest grand unified theory. However, this model 
is not realistic. In the second part of my talk 
I will present the minimal realistic extension of the 
Georgi-Glashow theory. I will show that this model 
could be tested at future proton decay experiments 
since the upper bound on the proton decay lifetime 
is $\tau_p \leq 1.4 \times 10^{36}$ years. This model 
predicts light scalar leptoquarks, therefore for the first time 
we have the possibility to test a GUT model at future 
hadron colliders, particularly at the LHC.

\section{Upper bound on the total proton lifetime}

There are several relevant operators to the decay of the proton. 
In non-supersymmetric grand unified theories we have the gauge 
and Higgs $d=6$ contributions. It is well-known that 
the gauge contributions are the most important in this context. 
In supersymmetric models the most important contributions are 
coming from the $d=4$ and $d=5$ operators (For details see~\cite{review}). 
As we know one can always forbid or suppress the $d=4$, $d=5$, and 
the Higgs $d=6$ contributions. However, it is not 
possible to forbid the gauge contributions in grand 
unified models with the standard model matter content. 
Now, since those contributions are always present 
and are the least model dependent, the upper bound on the proton 
lifetime is coming from the minimization of the 
decay rate due to the presence of those operators.     

In order to minimize the decay rate due to the presence of gauge 
$d=6$ contributions let us study those in details. In the physical 
basis these operators read as~\cite{PFP}:

\begin{eqnarray}
\label{Oec} 
\textit{O}(e_{\alpha}^C, d_{\beta})&=&
c(e^C_{\alpha}, d_{\beta}) \ \epsilon_{ijk} \ \overline{u^C_i} \
\gamma^{\mu} \ u_j \ \overline{e^C_{\alpha}} \
\gamma_{\mu} \ d_{k \beta}, \\
\label{Oe} 
\textit{O}(e_{\alpha}, d^C_{\beta})&=& c(e_{\alpha},
d^C_{\beta}) \ \epsilon_{ijk} \ \overline{u^C_i} \ \gamma^{\mu} \
u_j \ \overline{d^C_{k \beta}} \
\gamma_{\mu} \ e_{\alpha},\\
\label{On} 
\textit{O}(\nu_l, d_{\alpha}, d^C_{\beta} )&=& c(\nu_l,
d_{\alpha}, d^C_{\beta}) \ \epsilon_{ijk} \ \overline{u^C_i} \
\gamma^{\mu} \ d_{j \alpha}
\ \overline{d^C_{k \beta}} \ \gamma_{\mu} \ \nu_l 
\end{eqnarray} 
where
\begin{eqnarray}
c(e^C_{\alpha}, d_{\beta})&=& {k_1^2 \left[V^{11}_1 V^{\alpha \beta}_2 + ( V_1 V_{UD})^{1
\beta}( V_2 V^{\dagger}_{UD})^{\alpha 1}\right]}
\\
c(e_{\alpha}, d_{\beta}^C) &=& k^2_1  \ V^{11}_1 V^{\beta \alpha}_3 \ + \  k_2^2 \
(V_4 V^{\dagger}_{UD} )^{\beta 1} ( V_1 V_{UD} V_4^{\dagger} V_3)^{1 \alpha}  
\\
c(\nu_l, d_{\alpha}, d^C_{\beta})&=& k_1^2 \ ( V_1 V_{UD} )^{1 \alpha} (V_3 V_{EN})^{\beta l} \ 
+ \ k_2^2 \ V_4^{\beta \alpha}( V_1 V_{UD} V^{\dagger}_4 V_3 V_{EN})^{1l}
\end{eqnarray}
$\alpha=\beta=1,2$; $l=1,2,3$, and $k_{1,2}= g_{GUT} / \sqrt{2} {M_{V,V'}}$. 
In our notation $g_{GUT}$ is the gauge coupling at the GUT scale, and 
$M_{V,V'}$ the mass of the gauge bosons $(X, Y)=({\bf 3},{\bf 2},5/3)$, 
$(X', Y')=({\bf 3},{\bf 2},-1/3)$, respectively. The relevant mixing 
matrices for the above equations are defined by $V_1= U_C^{\dagger} U$, 
$V_2=E_C^{\dagger}D$, $V_3=D_C^{\dagger}E$, $V_4=D_C^{\dagger} D$, 
$V_{EN}=E^{\dagger}N=K_3 V_{PMNS}$, and $V_{UD}=U^{\dagger}D=K_1 V_{CKM} K_2$, 
once we define the diagonalization of the Yukawa matrices as: 
$U^T_C \ Y_U \ U = Y_U^{diag}$, $D^T_C \ Y_D \ D = Y_D^{diag}$, and 
$E^T_C \ Y_E \ E = Y_E^{diag}$. Here it is important to understand 
that it is difficult to predict the lifetime of the proton 
since we do not know all mixing matrices mentioned above. 
Even in minimal models it is very difficult to predict 
all those matrices. 

Now, we would like to find the minimum of the decay rate. 
Let us study the case of grand unified theories based 
on $SU(5)$ gauge symmetry ($k_2 = 0$). In order to 
find a possible ``minimum'' of the decay rate in these theories, 
we study two major cases~\cite{upper}. In the first case 
there are no decays into a meson and antineutrinos, 
and in the second there are no decays into a meson 
and charged antileptons. We investigated those cases 
in detail, showing that the upper bound is coming from 
the first case~\cite{upper}. Let us understand these results. 
In order to set to zero all channels with antineutrinos 
we have to look for a model of fermion masses where the 
following conditions are valid~\cite{upper}: 

\begin{eqnarray}
(V_1 V_{UD})^{1 \alpha} \ &=& (U_C^\dagger D)^{1 \alpha}\ =0 \Longrightarrow
U_C= D A^{\dagger} \ \ (\textrm{C.I})
\\
V_2^{\beta \alpha } \ &=& (E_C^\dagger D)^{\beta \alpha}\ =0
\Longrightarrow
E_C= D B_1 \ \ \ (\textrm{C.II}) 
\\
V_3^{\beta \alpha } \ &=& (D_C^\dagger E)^{\beta \alpha} \ =0
\ \Longrightarrow
D_C= E B_2 \ (\textrm{C.III}) 
\end{eqnarray}
with
\begin{equation}
A= \left( \begin{array} {ccc}
 0  & 0  &  e^{i\alpha} \\
 \ldots  & \ldots  & 0 \\
 \ldots  & \ldots  & 0 
\end{array} \right), 
\ B_{i} = \left( \begin{array} {ccc}
 0  & 0  &  e^{i\beta_{i}} \\
 0  & e^{i\gamma_{i}}  & 0 \\
 e^{i\delta_{i}}  & 0   & 0 
\end{array} \right) 
\end{equation}
In this case the decay rate is given by:
\begin{eqnarray}
\Gamma_p=\Gamma (p \to K^{0} \mu^+)&=& 
8 \pi^2 C(p,K^0) \ |V_{CKM}^{13}|^2 \ \alpha_{GUT}^2 
\ M_{V}^{-4}
\end{eqnarray}
where
\begin{displaymath}
C(p,K^0)= \frac{(m_p^2 - m_K^2)^2}{8 \pi m_p^3 f^2_{\pi}} \ A_L^2 \
|\beta|^2 \times \left[1 + \frac{m_p}{m_B} (D - F) \right]^2
\end{displaymath}
See reference~\cite{upper} for the notation. Therefore 
the upper bound on the proton lifetime reads as:
\begin{equation}
\tau_p \leq 6.0 \times 10^{39} \
\frac{(M_V/10^{16}\,\textrm{GeV})^4}{\alpha_{GUT}^2} \
(0.003\,\textrm{GeV}^3 / \beta)^2\,\textrm{years}
\end{equation}
Notice that this expression is valid for a given value 
of the GUT scale and $\alpha_{GUT}$. Now, using the 
experimental lower bound ($\tau_p \geq 50 \times 10^{32}$\,years) 
and the value $\beta=0.003$\,GeV$^3$ for the matrix element, 
we can conclude that the lower bound on the GUT scale 
is $M_V \ > \ 3.04 \times 10^{14} \sqrt{\alpha_{GUT}}\,\textrm{GeV}$.
Now, if we use the values for $\alpha_{GUT}=1/39-1/25$ in the non-susy 
and susy case, respectively:

\begin{equation}
{\label{GUT} M_V \ > \ (4.9-6.1) \times 10^{13} \,\textrm{GeV}}
\end{equation}

Notice that the lower bound on the GUT scale tells us that 
a non-SUSY $SU(5)$ could be consistent with 
the experimental bounds.

In flipped $SU(5)$ theories~\cite{Barr} ($k_1=0$) it 
is possible to set to zero all proton decay channels 
at the same time~\cite{rotate}. Therefore, the above 
upper bound on the proton lifetime is also valid 
for $SO(10)$ theories. 

Now, we can make a plot (Figure~1) to understand when a given 
grand unified theory is ruled out by the experimental 
lower bounds on the proton lifetime. 

\begin{figure}[h]
\begin{center}
\includegraphics[width=5in]{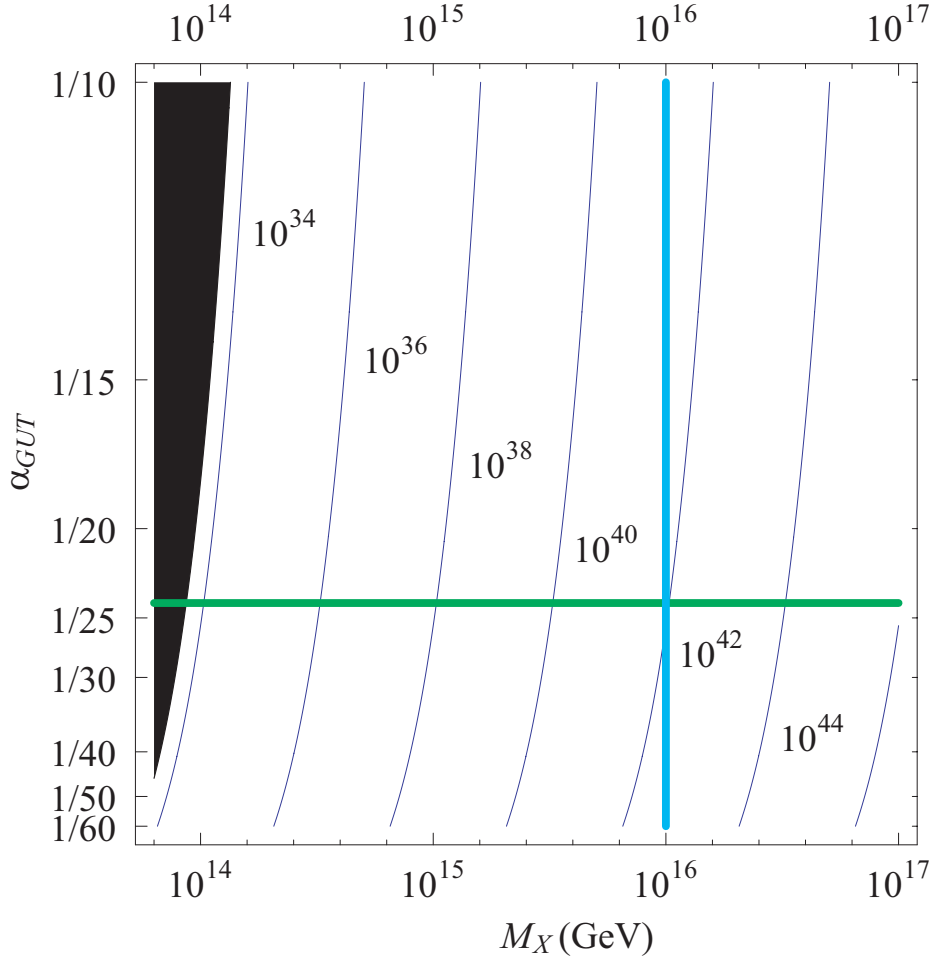}
\end{center}
\caption{\label{figure:1}
Isoplot for the upper bounds on the total proton 
lifetime in years in the Majorana neutrino case in the 
$M_X$--$\alpha_{GUT}$ plane. The value of the
unifying coupling constant is varied from $0.02$ to $1$~\cite{upper}.} 
\end{figure}

Notice that from Figure~1 we can conclude that the 
non-supersymmetric grand unified theories are still 
in agreement with the proton decay experiments. 
Now, since in future experiments the bounds on 
the proton decay experiments could be improved in two 
orders of magnitude, we are not sure about the possibility 
to test the supersymmetric grand unified theories through 
nucleon decay where the $d=5$ operators are absent 
or suppressed. In other words, in this case the 
lifetime of the proton could be very large.

As we discussed before the non-susy theories 
could be realistic. Then, let us study the 
possibility to write down the simplest non-susy 
grand unified theory.  

\section{The minimal realistic extension of the Georgi-Glashow model}

In the Georgi-Glashow model based on $SU(5)$ gauge symmetry~\cite{GG} 
the standard model matter is unified in the 
$\bar 5$ and $10$ representations and the Higgs sector is composed 
of the $5_H$ and $24_H$. As it is well-known 
this model is not realistic since it is not possible to 
unify all gauge couplings, the neutrinos are massless, and we find 
the wrong relation $Y_D=Y_E^T$. Now, as we pointed out 
in reference~\cite{MRSU5}, if we add the $15_H$ and use 
the higher-dimensional operators, we can write the minimal 
realistic extension of the Georgi-Glashow model, where it is 
possible to achieve unification, we can use the type II see-saw 
mechanism~\cite{seesaw} for neutrino masses and it 
is possible to find a consistent relation between the fermion 
masses in this context. 

We studied the unification constraints in detail~\cite{MRSU52}. 
In Figure~2 we show the whole parameter space where it is possible 
to achieve unification in agreement with all low energy data.

\begin{figure}[h]
\begin{center}
\includegraphics[width=5in]{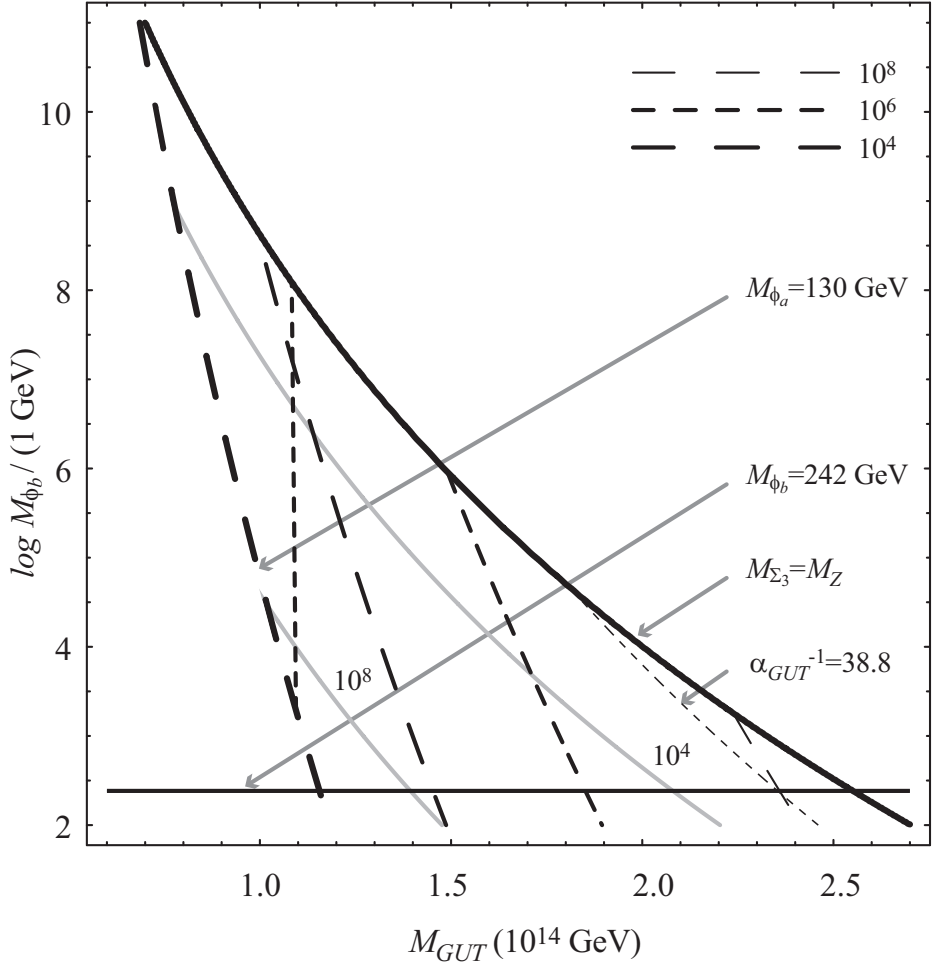}
\end{center}
\caption{\label{figure:2}Plot of lines of constant $M_{\Sigma_3}$ and
$M_{\Phi_a}$ in the $M_{GUT}$--$\textit{log}\, (M_{\Phi_b}/1\,\textrm{GeV})$
plane, assuming exact one-loop unification. We use the central values for the
gauge couplings as given in the text. All the masses are given in GeV units.
The triangular region is bounded from the left (below) by the experimental
limit on $M_{\Phi_a}$ ($M_{\Phi_b}$). The right bound is $M_{\Sigma_3} \geq
M_Z$. The two grey solid (thick dashed) lines are the lines of constant
$M_{\Sigma_3}$ ($M_{\Phi_a}$). The line of constant $\alpha^{-1}_{GUT}\,$ is
also shown. The region to the left of the vertical dashed line is excluded by
the proton decay experiments if $\beta=0.015$\,GeV$^3$~\cite{MRSU52}.}
\end{figure}     

Notice the different values for the masses of the fields which help 
us to achieve unification. Those are $\Sigma_3=(1,3,0) \subset 24_H$, 
$\Phi_a=(1,3,1) \subset 15_H$, and $\Phi_b=(3,2,1/6) \subset 15_H$. 
In this model the maximal value for the unification scale is 
$M_{GUT}=3.2 \times 10^{14}$ GeV, defined for $M_{\Sigma_8}=10^5$ GeV, 
$M_{\Sigma_3}=M_Z$, $M_{\Phi_a}=130$ GeV, $M_{\Phi_b}=242$ GeV, 
and $\alpha_{GUT}=1/37.3$. See references~\cite{MRSU5} and~\cite{MRSU52} 
for details. Now, using the maximal value for the GUT 
scale the upper bound on the proton lifetime reads as:
\begin{equation}
\tau_p  \ \leq 1.4  \ \times 10^{36}  \ \textrm{years}
\end{equation}
Therefore we can say that this model could be tested or 
ruled out at the next generation of proton decay experiments.

Recently, we studied the simplest renormalizable realistic 
$SU(5)$ model which Higgs sector is composed of $5_H$, 
$24_H$, and $45_H$~\cite{45H}, and we concluded that the proton 
lifetime can be very large. Therefore, we can say that 
the only realistic non-SUSY $SU(5)$ that we can 
verify in near future is the model presented 
in reference~\cite{MRSU5}.

Now, if we study carefully the results presented in Figure~2, we see 
that once we impose the proton decay constraints we get an 
upper bound on the scalar leptoquark mass, which is basically 
$M_{\Phi_b} \leq 10^8$ GeV. Also notice that in the case 
of the most natural implementation of the type II 
see-saw mechanism (large $M_{\Phi_a}$) the mass of the scalar leptoquarks 
($\Phi_b$) is in the phenomenologically interesting region 
$(O(10^2-10^3) GeV)$. Then, we can conclude that our scenario 
has a potential to be tested at the next generation of 
collider experiments, particularly at the Large Hadron 
Collider (LHC) at CERN. See reference~\cite{leptoquarksLHC} 
for the study of leptoquarks at the LHC.   

Notice that usually it is very difficult to be sure about 
the possibility to test any grand unified theory 
through proton decay. However, here we have the exciting 
possibility to verify the idea of the unification of the 
gauge interactions at future colliders.  

\section{Summary}

In this talk I have discussed the possibility to find an 
upper bound on the proton decay lifetime. The conservative 
upper bound on the total proton decay valid for any 
grand unified model with the standard model 
matter content is given by:
   
\begin{equation}
\tau_p \leq 6.0 \times 10^{39} \
\frac{(M_V/10^{16}\,\textrm{GeV})^4}{\alpha_{GUT}^2} \
(0.003\,\textrm{GeV}^3 / \beta)^2\,\textrm{years}
\end{equation}
where $M_V$ is the mass of the superheavy gauge bosons 
and $\beta$ is the value of the matrix element.
 
We conclude that the non-supersymmetric 
grand unified theories with low unification scale 
$(M_V \ > \ 4.9 \times 10^{13} \,\textrm{GeV})$ 
are not ruled out by the bounds coming from the 
proton decay searches. 

It is shown that the minimal non-supersymmetric 
grand unified theory is the modified 
Georgi-Glashow model with Higgs sector composed 
of $5_H$, $24_H$, and $15_H$. This scenario can be 
tested at future proton decay experiments 
($\tau_p  \ \leq 1.4  \ \times 10^{36}  \ \textrm{years}$) 
and future colliders, particularly at LHC, 
through the production of scalar leptoquarks $\Phi_b$. 

\begin{acknowledgments}
{\small I would like to thank the organizers of 
NURT'06 in Havana, PLANCK'06 in Paris, and SUSY'06 in Irvine 
for the possibility to present my work and for those stimulating 
conferences. This work has been supported 
by {\em Funda\c{c}\~{a}o para a Ci\^{e}ncia e a
Tecnologia} (FCT, Portugal) through the project CFTP, POCTI-SFA-2-777 and a
fellowship under project POCTI/FNU/44409/2002. 
I would like to thank I.~Dorsner and R.~Gonz\'alez Felipe 
for many discussions and enjoyable collaboration. I would like to thank 
B. Bajc, F. Feruglio, M. Frigerio, W. Hollik, S. Lavignac, P. Nath, 
M. Nebot, G. Rodrigo Garcia, C. Savoy, and G. Senjanovi\'c for discussions 
and comments.}  
\end{acknowledgments} 


\end{document}